\documentclass[a4paper,twocolumn,
english,aps,prl,floatfix]{revtex4}
\usepackage[latin1]{inputenc}
\usepackage{amsmath}
\usepackage{babel}
\usepackage{graphics}
\usepackage{amssymb}



\begin{document}

\newcommand{\dt}{\Delta\tau}
\newcommand{\al}{\alpha}
\newcommand{\ep}{\varepsilon}
\newcommand{\ave}[1]{\langle #1\rangle}
\newcommand{\have}[1]{\langle #1\rangle_{\{s\}}}
\newcommand{\bave}[1]{\big\langle #1\big\rangle}
\newcommand{\Bave}[1]{\Big\langle #1\Big\rangle}
\newcommand{\dave}[1]{\langle\langle #1\rangle\rangle}
\newcommand{\bigdave}[1]{\big\langle\big\langle #1\big\rangle\big\rangle}
\newcommand{\Bigdave}[1]{\Big\langle\Big\langle #1\Big\rangle\Big\rangle}
\newcommand{\braket}[2]{\langle #1|#2\rangle}
\newcommand{\up}{\uparrow}
\newcommand{\dn}{\downarrow}
\newcommand{\bb}{\mathsf{B}}
\newcommand{\ctr}{{\text{\Large${\mathcal T}r$}}}
\newcommand{\sctr}{{\mathcal{T}}\!r \,}
\newcommand{\btr}{\underset{\{s\}}{\text{\Large\rm Tr}}}
\newcommand{\lvec}[1]{\mathbf{#1}}
\newcommand{\gt}{\tilde{g}}
\newcommand{\ggt}{\tilde{G}}
\newcommand{\jpsj}{J.\ Phys.\ Soc.\ Japan\ }
\newcommand{\rf}[1]{(\ref{#1})}
\newcommand{\cao}{\c c\~ao\ }
\newcommand{\coes}{\c c\~oes\ }
\newcommand{\jpcm}{J.Phys.: Condens.\ Matter\ }
\newcommand{\jap}{J.\ Appl.\ Phys.\ }

\title{Critical temperature for the two-dimensional attractive 
Hubbard Model} 

\author{Thereza Paiva$^{\, (1)}\!$,  Raimundo R.\ dos Santos$^{\, (1)}\!$, 
R.\ T.\ Scalettar$^{\, (2)}\!$, and P.\ J.\ H.\ Denteneer$^{\, (3)}\!$ }

\affiliation{$^{(1)}$Instituto de F\' \i sica, 
		Universidade Federal do Rio de Janeiro,
                Cx.P.\ 68.528, 
		21945-970 Rio de Janeiro RJ, Brazil\\
	$^{(2)}$Department of Physics, 
		University of California, Davis,
		California 95616-8677 \\
	$^{(3)}$Instituut-Lorentz, Leiden University,
		P.O. Box 9506, 
		2300 RA Leiden, The Netherlands}



\date{Version 11.01 -- \today}

\begin{abstract}
The critical temperature for the attractive Hubbard model on a square 
lattice is determined from the analysis of two independent quantities, 
the helicity modulus, $\rho_s$, and the pairing correlation 
function, $P_s$. These quantities have been calculated through Quantum 
Monte Carlo simulations for lattices up to $18\times 18$, and for several 
densities, in the intermediate-coupling regime. Imposing the universal-jump 
condition for an accurately calculated $\rho_s$, together with thorough 
finite-size scaling analyses (in the spirit of the phenomenological 
renormalization group) of $P_s$, suggests  that $T_c$ is 
considerably higher than hitherto assumed.
\end{abstract}

\maketitle

The attractive Hubbard model\cite{MRR90,Wilson01}
has been successfully used to elucidate
a number of important and fundamental issues in both
conventional and high-temperature (cuprate) superconductivity.
The nature of the crossover between BCS superconductivity (at weak
coupling, or small on-site attraction) and Bose-Einstein condensation of
tightly bound pairs (strong coupling) has been shown to be
smooth \cite{Leggett80,Randeria98}.  The appearance of
preformed pairs within a certain range of parameters in the normal phase,
especially below a characteristic temperature, has been related to
pseudogap behavior of high-temperature
superconductors \cite{Randeria92,dS94}.  
Further, this model allows one to introduce disorder on the fermionic
degrees of freedom \cite{Litak92,Scalettar99} and investigate the behavior
near the quantum critical point of the two-dimensional
insulator-superconductor transition; this provides an alternative to the
dirty-boson picture \cite{Fisher90} to discuss the universal conductivity 
\cite{Hsu95}. 
The attractive Hubbard model with  a periodic modulation of $U$ has been 
used to interpret superconductivity in layered structures \cite{Paiva}.

A basic concern has run through many of these 
calculations, in particular those based on Quantum Monte Carlo (QMC)
simulations.
In two dimensions, there is a consensus that the
early QMC phase diagram \cite{Scalettar89,Moreo91} --
in the space of critical temperature, $T_c$, electronic density, $\langle
n \rangle$, and magnitude of the on-site attraction, $|U|$ -- is
qualitatively correct. However, some serious quantitative discrepancies
have emerged over the years, pointing towards higher critical
temperatures; see, e.g., the Bogoliubov-Hartree-Fock (BHF) approach of 
Ref.\
\onlinecite{Denteneer91-93}. Our purpose here is to examine the
dependence of $T_c$ with $\langle n\rangle$, for fixed $U$, by resorting
to a much wider (namely, larger system sizes and several electronic
densities) set of QMC data, together with alternative procedures to locate
the critical temperature.
Establishing an accurate value for this most fundamental property of the model
is important, especially as the physics of variants of the attractive
Hubbard Hamiltonian is explored, and comparisons are made to the original 
system.

The model is defined by the Hamiltonian
\begin{eqnarray}
&{\cal H}=&-t\sum_{\langle{\bf i},{\bf j}\rangle,\,\sigma}
 \left(c_{{\bf i}\sigma}^{\dagger}c_{{\bf j}\sigma}^{\phantom{\dagger}}+ 
{\rm
H.c.}\right) 
-\mu \sum_{\bf i}
\left(n_{{\bf i}\uparrow}^{\phantom{\dagger}}+
      n_{{\bf i}\downarrow}^{\phantom{\dagger}}\right)
\nonumber\\
&&
-|U|\sum_{\bf i} \left(n_{{\bf i}\uparrow}^{\phantom{\dagger}}-
         \frac{1}{2}\right)
         \left(n_{{\bf 
i}\downarrow}^{\phantom{\dagger}}-\frac{1}{2}\right)
\label{Hub}
\end{eqnarray}
where 
$c_{{\bf i}\sigma}^{\phantom{^\dagger}}$ 
($c_{{\bf i}\sigma}^{^{\dagger}}$) 
destroys (creates) an electron with spin $\sigma$ on site ${\bf i}$ of a
square lattice, $\langle {\bf i},{\bf j} \rangle $ denotes nearest 
neighbor sites,
$n_{{\bf i}\sigma}\equiv c_{{\bf i}\sigma}^{^{\dagger}}c_{{\bf
i}\sigma}^{\phantom{^\dagger}}$, $|U|$ is the strength of the attractive
interaction and $\mu$ is the chemical potential. From now on, all energies 
are expressed in units of the hopping amplitude, $t$, and we also set 
$k_B=1$.

At half filling (which corresponds to the particle-hole symmetric point,
$\mu=0$), the degeneracy of charge-density wave (CDW) and singlet
superconducting (SS) correlations leads to a three-component order
parameter \cite{shiba}; the transition temperature is therefore suppressed
to zero. Away from half-filling, CDW correlations are suppressed and a
finite temperature Kosterlitz-Thouless (KT) transition \cite{kt} into a SS
phase takes place \cite{Moreo91,Assaad94}; this phase has only
algebraically decaying correlations for $0 < T \leq T_c $. Further, close
to half filling an exact mapping onto the two-dimensional Heisenberg
antiferromagnetic model in a magnetic field leads
to \cite{Scalettar89,Moreo91} $T_c \simeq -2\pi J/ \ln |1-\langle n
\rangle| $, so that $T_c$ rises sharply from zero as one dopes away from
$\langle n \rangle=1$.

We start by employing the analysis of Ref.\ \onlinecite{Moreo91}
to new data for the SS pairing correlation function,
\begin{equation}
P_s= \langle \Delta^\dagger \Delta + \Delta \Delta^\dagger \rangle,
\label{Ps-def}
\end{equation}
with
\begin{equation}
\Delta^\dagger = \frac{1}{\sqrt {N}} \sum_i
c_{{\bf i}\uparrow}^{^\dagger}c_{{\bf i}\downarrow}^{^\dagger}.
\label{Delta}
\end{equation}
For $ 0 < T \leq T_c $, one expects 
\begin{equation}
\Gamma(r)\equiv \langle c_{{\bf i}\uparrow}^\dagger c_{{\bf 
i}\downarrow}^\dagger
c_{{\bf j}\downarrow}^{\phantom{\dagger}}
c_{{\bf j}\uparrow}^{\phantom{\dagger}} + {\rm H.c.}\rangle
\sim r^{-\eta(T)}
\label{Gr}
\end{equation}
where $r\equiv |{\bf i}-{\bf j}|$, and  $\eta(T)$ 
increases monotonically between $\eta(0)=0$ and 
$\eta(T_c)=1/4$ \cite{kt,Berche02}.  

\begin{figure}
{\centering\resizebox*{3.3in}{!}{\includegraphics*{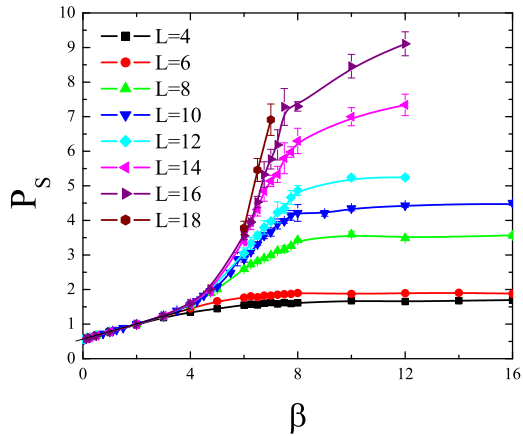}}}
\caption{(Color online) $P_s$ as a function of $\beta\equiv 1/T$ for 
$\langle n\rangle =
0.5$ and different lattice sizes, $L$.}
\label{n05raw} 
\end{figure}

The finite-size scaling behavior of $P_s$ is therefore obtained upon 
integration of $\Gamma(r)$ over a two-dimensional system of \emph{linear} 
dimension $L$. One then has \cite{Moreo91}
\begin{equation}
P_s=L^{2 - \eta(T_c)} f(L/ \xi), \ \ \ L\gg 1,\ T\to T_c^+, 
\label{Ps}
\end{equation}
with 
\begin{equation}
\xi \sim \exp \left[ \frac{A}{(T-T_c)^{1/2}} \right];
\label{xi}
\end{equation}
in the thermodynamic limit, one recovers $P_s\sim\xi^{7/4}$. 
For completeness, one should mention that since $\eta\to 0$ as $T\to 0$, 
the system displays long-range order in the ground state, so that a 
`spin-wave scaling' is expected to hold \cite{Huse88}:
\begin{equation}
\frac{P_s}{L^2}= |\Delta_0|^2 + \frac{C}{L},
\label{Ps-sw}
\end{equation}
where $\Delta_0$ is the superconducting gap function at zero 
temperature, and $C$ is a $|U|$-dependent constant.

Similarly to Ref.\ \onlinecite{Moreo91}, here we use the determinant
QMC algorithm \cite{qmc} to calculate $P_s$.  Typically our
data have been obtained after 500 warming-up steps followed by 50,000
sweeps through the lattice. The discretized imaginary-time
interval \cite{qmc} was set to $\Delta\tau=0.125$, which is small
enough for the results not to depend on this choice in any significant
way.

\begin{figure}
{\centering\resizebox*{3.3in}{!}{\includegraphics*{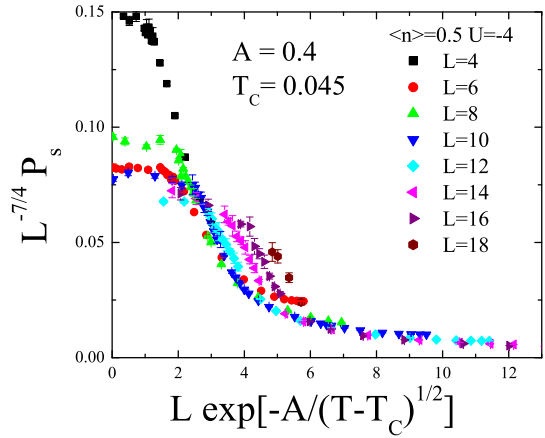}}}
\caption{(Color online) Rescaled $P_s$ as a function of $w\equiv L \exp
[-A/(T-T_c)^{1/2}]$ for $\langle n\rangle = 0.5$ and different lattice
sizes, $L$.  The values for $A$ and $T_c$ are the ones 
determined in Ref.\ \onlinecite{Moreo91}.}
\label{n05rescaleA} 
\end{figure}

In Fig.\ \ref{n05raw} we show raw data for $P_s$ as a function of $(1/T)$, for fixed density,
$\langle n \rangle =0.5$ and $U=-4$, and for different lattice sizes. 
The crossover between temperature- and size-limited regimes is 
described by finite-size scaling (FSS) theory \cite{Cardy} and appears as 
a levelling off of $P_s$ below a certain temperature for each system size.
Before a more quantitative scaling analysis, we can already see a 
suggestion that $T_c$ is around 1/6 from the raw $P_s$ data.  In general,
at temperatures for which correlations are short ranged, a structure 
factor like $P_s$ is independent of lattice size.  As $T$ is decreased,  
the point at which the structure  factor begins growing with lattice size 
signals the temperature at which the correlation length $\xi$ is becoming 
large (comparable to the lattice size $L$), thus providing a crude 
estimate of $T_c$.  The subsequent plateau  at low temperatures occurs 
when $\xi >> L$. 
This crossover is contemplated by the FSS form, Eq. (5), which can be 
invoked to determine $T_c$ by plotting $L^{-7/4}P_s$ as a function of
of $w\equiv L\exp [-A/(T-T_c)^{1/2}]$,
at a given $U$, for different system sizes, with $T_c$ and $A$ being
adjusted to give the best possible data collapse, as done in Ref.\
\onlinecite{Moreo91}. Figure \ref{n05rescaleA} shows the resulting scaling 
plot, in which
the values $A=0.4$ and $T_c=0.045$ were determined in Ref.\
\onlinecite{Moreo91}.  
With our substantially increased amount of data points it becomes clear 
that the data collapse onto a single curve with the parameters $A$ and
$T_c$ of Ref.\ \onlinecite{Moreo91} becomes rather unsatisfactory.
We furthermore note that Eq.\ (\ref{xi}) is expected to hold only for
$t\lesssim 10^{-2}$ ($t \equiv (T-T_c)/T_c$), see, e.g., Ref.\
\onlinecite{gupta}. For the value of  $T_c$ used in 
Fig.\ \ref{n05rescaleA}, only few data obtained for 
$L=4, 6, 8, 10$ in Ref.\ \onlinecite{Moreo91} satisfy this criterion.

\begin{figure}
{\centering\resizebox*{3.3in}{!}{\includegraphics*{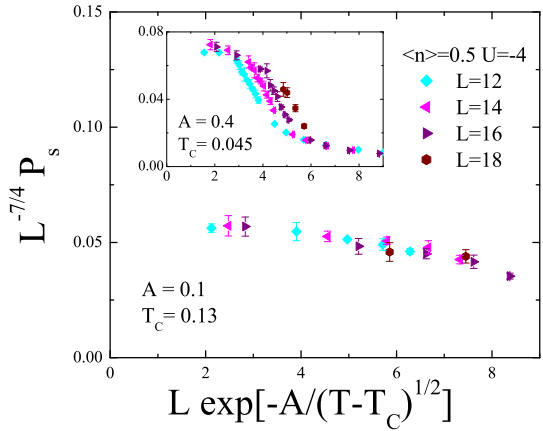}}}
\caption{(Color online) Same as Fig.\ \ref{n05rescaleA}, but with $A$ and 
$T_c$ 
determined from the present data. Inset shows same system sizes, with $A$ 
and $T_c$ from Ref.\ \onlinecite{Moreo91}.}
\label{n05rescale1} 
\end{figure}

We therefore obtain new values of $A$ and $T_c$ from our expanded data set.
We disregard the data from the smallest system sizes, $L=4$ (which, 
additionally, has a special topology, being equivalent to a $2\times 
2\times 2 \times 2$ four-dimensional lattice), $L=6$, $L=8$ and also 
$L=10$; as we will see below, the SS pairing correlation function 
presents large finite size effects for these lattice sizes. 
Furthermore, we only include data points for temperatures $T$ for which
Eq.\ (\ref{xi}) is expected to hold (see above). 
Fig.\ \ref{n05rescale1} clearly shows that, for the larger latices, 
our newly determined parameters $A=0.1$ and critical temperature $T_c=0.13$ 
render a much better data collapse than the old parameters.

The present analysis shows that the estimates of $T_c$ obtained in Ref.\
\onlinecite{Moreo91} can be quantitatively quite unreliable. In our opinion,
this is due to the fact that the finite-size behaviour of $P_s$,
Eqs.\ (\ref{Ps})-(\ref{xi}), follows from an analysis which is valid 
only for large enough lattice sizes, 
since it involves the binding-unbinding of rather
large structures (vortices) in the KT transition \cite{kt}.
Moreover, the parameters $A$ and $T_c$ that have to be found via data fitting
both reside in an exponent, resulting in large uncertainties for the
individual fitted parameters. Although the lattice sizes used in the
present study may not be large enough to determine $T_c$ with high
accuracy, our result is bound to be an improvement and in any case
indicates that the actual $T_c$ may be much larger (by even a factor of 3)
than believed sofar.

This tendency towards higher critical temperatures appears 
as well in a completely
independent analysis, based on the behavior of the helicity modulus (HM). 
The latter is a measure of the response of the system in the ordered phase 
to a `twist' of the order parameter \cite{Fisher73}, and can be expressed 
in terms of the current-current correlation functions as follows 
\cite{Scalapino92}
\begin{equation}
\rho_s=\frac{D_s}{4\pi e^2}=
\frac{1}{4} [\Lambda^L - \Lambda^T],
\label{Ds}
\end{equation}
where $D_s$ is the superfluid weight, and
\begin{equation}
\Lambda^L \equiv \lim_{q_x \to 0} \Lambda _{xx} (q_x, q_y=0,\omega_n=0),
\end{equation}
and
\begin{equation}
\Lambda^T \equiv \lim_{q_y \to 0} \Lambda _{xx} (q_x=0, q_y,\omega_n=0),
\end{equation}
are, respectively, the limiting longitudinal and transverse responses, 
with
\begin{equation}
\Lambda_{xx}(\vec q, \omega_n)= \sum_{\vec \ell} \int_0^\beta d\tau\ e^{i
\vec{q} \cdot \vec {\ell}} e^{i \omega_n \tau} \Lambda_{xx}(\vec \ell,
\tau),
\label{lambdaq}
\end{equation}
where $\omega_n=2 n \pi T$; 
\begin{equation}
\label{lambda}
\Lambda_{xx}(\vec \ell, \tau)= \langle j_x( \vec \ell, \tau) j_x (0,0) \rangle,
\end{equation}
where
\begin{equation}
j_x(\vec \ell,\tau)= e^{ {\cal H} \tau} [ it\sum_\sigma ( c_{\vec \ell + \hat x,\sigma}^{^{\dagger}}
c_{\vec \ell,\sigma} -  c_{\vec \ell,\sigma}^{^{\dagger}} c_{\vec \ell+ \hat x,\sigma} ] e ^{-{\cal H} \tau}
\end{equation}
is the $x$-component of the current density operator; see Ref.\
\onlinecite{Scalapino92} for details.

\begin{figure}
{\centering\resizebox*{3.3in}{!}{\includegraphics*{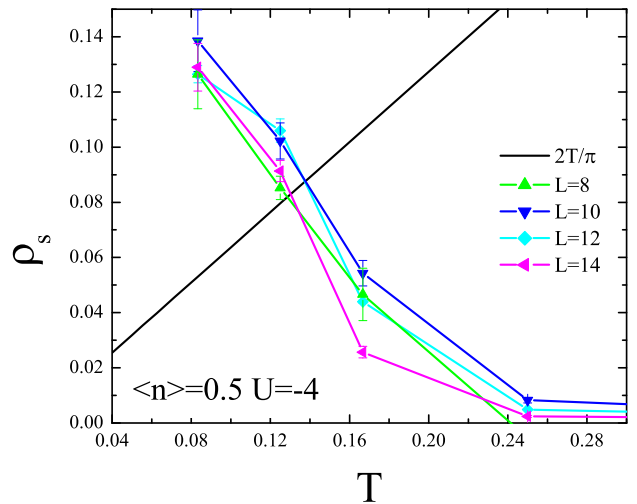}}}
\caption{(Color online) Helicity modulus as a function of temperature for 
$\langle
n\rangle = 0.5$ and different lattice sizes, $L$. The
straight line corresponds to $2T/\pi$.}
\label{rhosn05} 
\end{figure}

At the KT transition, the following universal-jump relation involving the 
helicity modulus holds \cite{Nelson77}:
\begin{equation}
T_c = \frac {\pi} {2} \rho_s^-,
\end{equation}
where $\rho_s^-$ is the value of the helicity modulus just below the
critical temperature. Thus we can obtain $T_c$ by plotting $\rho_s(T)$,
and looking for the intercept with $2T/\pi$. This procedure has been used
before, with $\rho_s$ calculated within a Bogoliubov-Hartree-Fock (BHF)
approximation \cite{Denteneer91-93,Denteneer94}; since transverse
current-current correlations were neglected, $\rho_s$ is likely to have 
been overestimated, and the ensuing $T_c$'s may have been too high. Here 
we 
calculate
both $\Lambda^L$ \emph{and} $\Lambda^T$ by QMC simulations to obtain
$\rho_s$ through Eq.\ (\ref{Ds}); a typical example of $\rho_s(T)$, for
$\langle n\rangle = 0.5$, is shown in Fig.\ \ref{rhosn05}.  We see that
finite-size effects are not too drastic, since all curves cross the
straight line within a small range of temperatures; that is, from Fig.\
\ref{rhosn05} we can estimate $T_c = 0.14 \pm 0.02$.

\begin{figure}
{\centering\resizebox*{3.3in}{!}{\includegraphics*{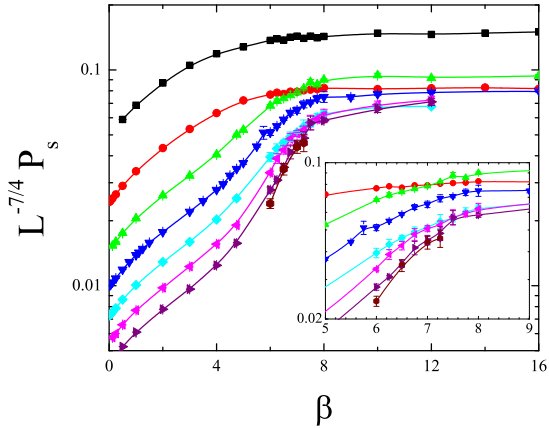}}}
\caption{(Color online) Logarithm-linear plot for the rescaled $P_s$ as a 
function 
of $\beta$, for $\langle n\rangle = 0.5$ and for different lattice 
sizes, $L$, symbols are the same as in Fig.\ 2. The inset shows a blow-up 
of the region centered about 
$\beta=7$. No parameters are adjusted.}
\label{betan05} 
\end{figure}

In order to check the robustness of this method, we can extract $T_c$ from
$P_s$ through a `phenomenological renormalization group' (PRG)
\cite{Nightingale82,dS81} analysis, provided some subtleties peculiar to
the KT transition are kept in mind. Since $\xi\to\infty$ for all $T< T_c$,
Eq.\ (\ref{Ps}) implies that curves for $L^{-7/4}P_s(L,\beta)$, when
plotted as functions of $\beta$, and for different $L$, should all
\emph{merge} for $\beta>\beta_c$. Figure \ref{betan05} shows that this
characteristic feature only sets in for the largest system sizes, namely,
$L\geq 12$, from which we can infer $\beta_c=7.5 \pm 0.25$; these error
bars are somewhat arbitrary, and result from visual inspection. It should
be stressed that this estimate for $\beta_c$ agrees remarkably well with
the one obtained from the helicity modulus, indicating the robustness of
both procedures to extract $T_c$. Interestingly, we should notice that the
curves for $L=6$ and 8 \emph{cross} each other (as in an ordinary second
order transition) at $\beta\simeq7$, which is very close to $\beta_c$
estimated from the larger systems.  Therefore, within the context of PRG,
for the smallest sizes a KT transition appears as an ordinary transition,
only crossing over to the merging feature for the largest sizes.

\begin{figure}
{\centering\resizebox*{3.3in}{!}{\includegraphics*{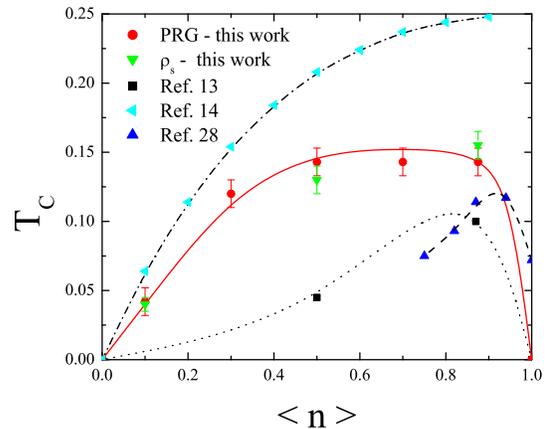}}}
\caption{(Color online) Critical temperature as a function of 
band-filling, obtained by 
different methods. All lines through data points 
are guides to the eye.}
\label{diag} 
\end{figure}

The critical temperature has been estimated for other electronic
densities, $\ave{n}=0.1$ (HM and PRG), 0.3 (PRG), 0.7 (PRG), and 0.875 (HM
and PRG); all PRG plots display the crossing and merging tendency observed
for $\ave{n}=0.5$. The resulting phase diagram is shown in Fig.\
\ref{diag}; for comparison, we also plot the early QMC results
\cite{Moreo91}, the parquet data from Luo and Bickers \cite{Luo93}, and
the estimates from the BHF approximation \cite{Denteneer91-93}. While
close to half filling all results (but BHF) are in fair agreement, for
larger dopings agreement is only found between the results from PRG and
those from the helicity modulus. The inescapable conclusion is that the
critical temperature for the superconducting transition in the attractive
Hubbard model is actually higher than previously assumed.

In summary, we have established very reliable estimates for the critical
temperature of the square-lattice attractive Hubbard model.  This is done
by finding quantitative agreement between entirely different procedures to
extract $T_c$ from two independent correlation functions (both computed by
determinant QMC).  As a result, the critical temperature is found to be
substantially higher than the currently accepted values, also obtained
using QMC, but with a different data analysis; as expected,
they are also substantially lower than estimates obtained within 
a Hartree-Fock/mean field approximation.

\begin{acknowledgments} 

The authors are grateful to S.\ de Queiroz and A.\ Moreo for discussions.  
TP and RRdS acknowledge partial financial support by Brazilian Agencies
FAPERJ, CNPq, Instituto do Mil\^enio para Nanoci\^encias/MCT, and Rede
Nacional de Nanoci\^encias/CNPq; RTS acknowledges support by
NSF-DMR-0312261.  This research is further supported by a joint
CNPq-690006/02-0/NSF-INT-0203837 grant.

\end{acknowledgments}

\end{document}